\documentclass[review,number,sort&compress,american]{elsarticle}
\usepackage{hyperref}
\usepackage{lineno}
\usepackage{graphicx}
\usepackage{multirow}
\usepackage{xfrac}
\usepackage[utf8]{inputenc}
\usepackage{amsmath}
\usepackage{amssymb}

\begin{document}

\begin{frontmatter}
\title{Electromagnetic calorimeter time measurement applications in the SND physics analysis}

\author[binp]{N A Melnikova\corref{mycorrespondingauthor}}
\cortext[mycorrespondingauthor]{Corresponding author}
\ead{n.a.melnikova@inp.nsk.su}
\author[binp,nsu]{M N Achasov}
\author[binp]{A A Botov}
\author[binp,nsu]{V P Druzhinin}
\author[binp,nsu]{L V Kardapoltsev}
\author[binp,nsu]{A A Korol}
\author[binp]{D P Kovrizhin}
\author[binp,nsu]{S I Serednyakov}
\author[binp,nsu]{I K Surin}

\address[binp]{Budker Institute of Nuclear Physics,\\11, Acad. Lavrentieva Pr., Novosibirsk, 630090 Russian Federation}
\address[nsu]{Novosibirsk State University,\\1, Pirogova str., Novosibirsk, 630090, Russian Federation}

\begin{abstract}
The SND is a non-magnetic detector at the VEPP-2000 $e^{+} e^{-}$ collider (BINP, Novosibirsk) designed for hadronic cross-section measurements in the center-of-mass energy range up to $2$ GeV. The important part of the detector is a hodoscopic electromagnetic calorimeter (EMC) with three layers of NaI(Tl) counters. The EMC signal shaping and digitizing electronics based on FADC allow to obtain both the signal amplitude and the arrival time. We describe the EMC signal processing and how the EMC measured time is applied in event reconstruction and physics analysis.
\end{abstract}


\end{frontmatter}

\section{Introduction}
\label{intro}
The SND \cite{snd1,snd2}  is a general purpose non-magnetic detector at the $e^{+} e^{-}$ collider VEPP-2000 \cite{vepp2k} designed for studying processes of $e^{+} e^{-}$ annihilation into hadrons up to $2$~GeV. 

\begin{figure*}[ht]
	\centering
	\includegraphics[width=0.8\columnwidth]{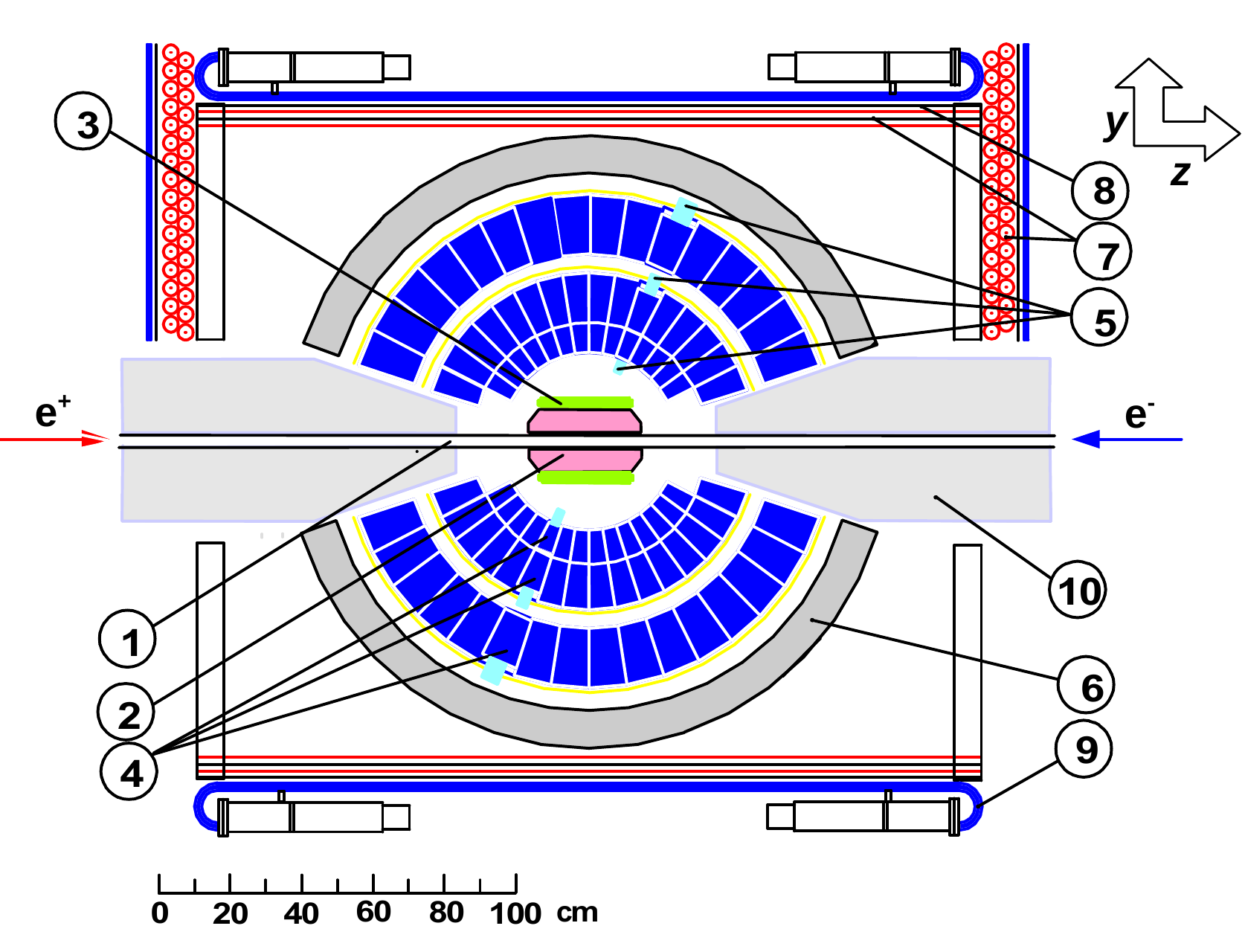}
	\protect\caption{The SND scheme \cite{snd2}: 1 \textemdash{} vacuum pipe, 2 \textemdash{} tracking system (TS), 3 \textemdash{} threshold Cherenkov counter, 4\textendash 5 \textemdash{} electromagnetic calorimeter (NaI (Tl)) (EMC), 6 \textemdash{} iron absorber, 7\textendash 9 \textemdash{} muon detector, 10 \textemdash{} focusing solenoids.}
	\label{fig:SNDScheme:}
\end{figure*}

The SND has several subsystems (Fig.~\ref{fig:SNDScheme:}): a cylindrical tracking system, an electromagnetic calorimeter (EMC), threshold Cherenkov counters and a muon detector. 

The EMC is the  main part of the SND. It provides uniform particle detection in a solid angle of $0.95 \cdot 4\pi$. The EMC consists of three layers of counters based on  NaI(Tl) crystals.

The tracking system (TS) \cite{TSSND2K} consists of a 9-layer  drift chamber (DC) with 24 jet cells and a proportional chamber in a common gas volume. 

Since 2018 the EMC spectrometric channel with a Flash ADC provides digitized signal pulses, which are processed to extract energy and time data \cite{newChannel_nim2016}. The main purpose of this work is to describe the time measurement process and the use of the EMC measured time in event reconstruction and physics analysis.

\section{Spectrometric channel}
\label{sec:channel}
\begin{figure}[ht]
	\centering
	\includegraphics[width=0.8\columnwidth]{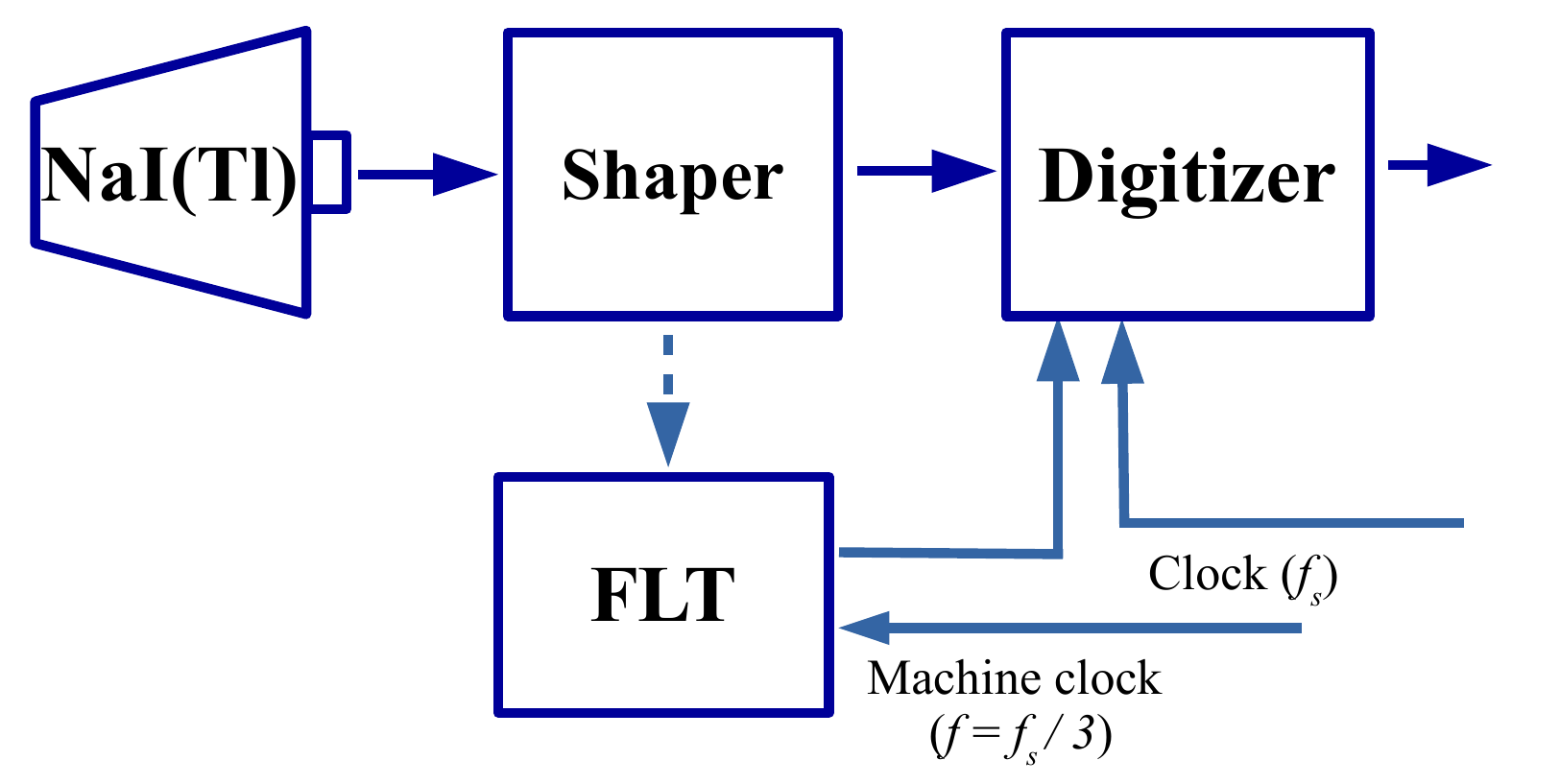}
	\protect\caption{The schematic layout of the EMC spectrometric channel. The abbreviations are described in the text.}
	\label{fig:emc_channel:}
\end{figure}
The schematic layout of the EMC spectrometric channel \cite{newChannel_nim2016} is shown in Fig.~\ref{fig:emc_channel:}. The channel includes a NaI(Tl) crystal with a vacuum phototriod as a readout device, a charge-sensitive preamplifier attached to it, a shaping amplifier (Shaper), and a digitizing module (Digitizer) with 12-bit flash analog-to-digital converters (FADCs).  Shapers also provide analog signals for a first-level trigger (FLT) signal. Shaped signals go through Digitizers where the sampling is performed at the frequency  $f_\mathrm{s}  \approx 36.9$~MHz and with the sampling period $T_\mathrm{s} \approx 27$~ns. The digitized pulses are read out after the arrival of the FLT signal, which is synchronized with a VEPP-2000 machine clock at the beam crossing frequency $f =\sfrac{f_\mathrm{s}}{3} \approx 12.3$~MHz (time between collisions $T_\mathrm{b} = 3 \,T_\mathrm{s} \approx 81$~ns). 

\subsection{Signal properties}
\label{subsec:signal}
\begin{figure}[ht]
	\centering
	\includegraphics[width=0.9\columnwidth]{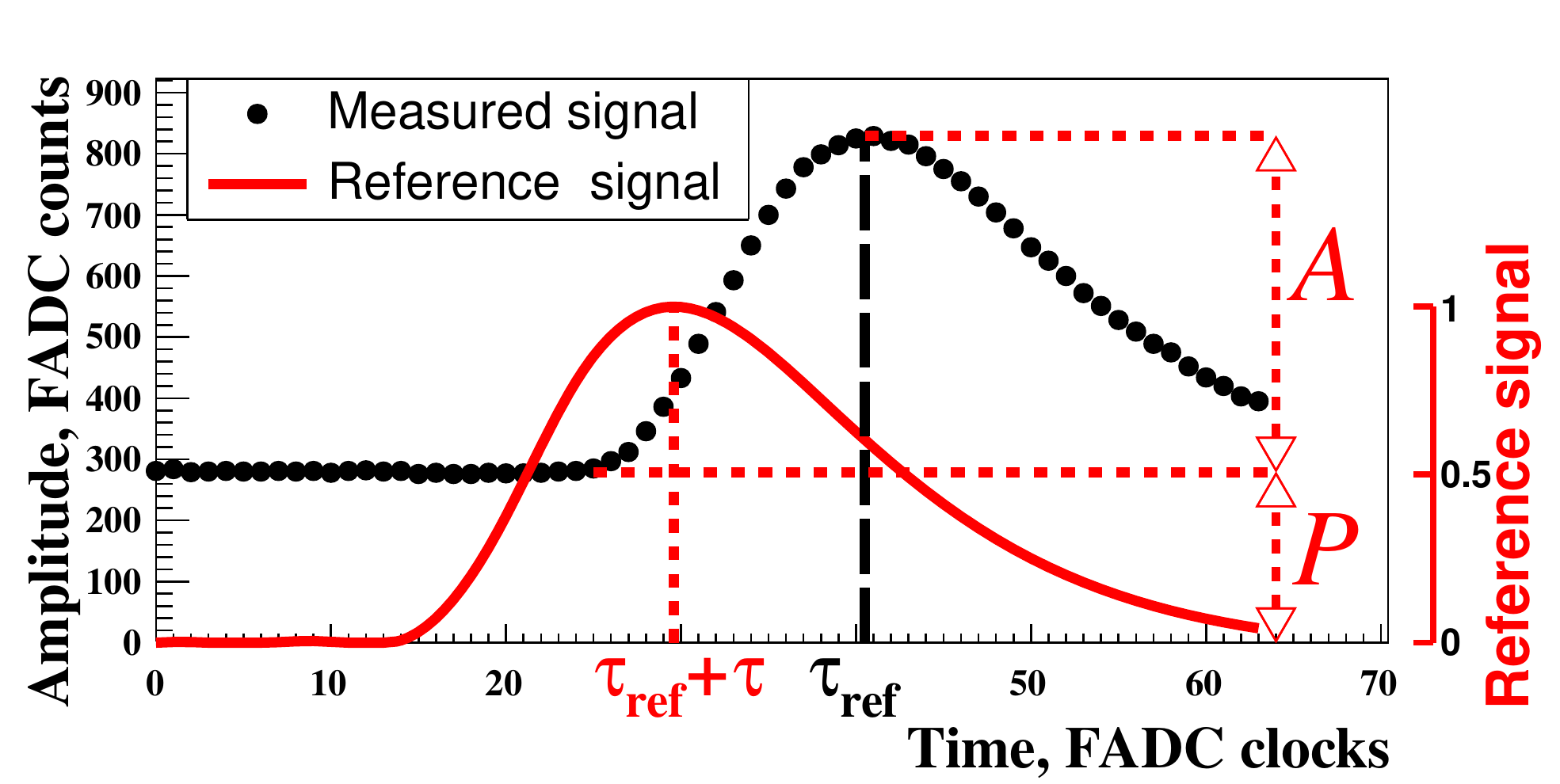}
	\protect\caption{EMC measured and reference signals. The abbreviations are described in the text.}
	\label{fig:emc_signal_params:}
\end{figure}

A  digitized  signal pulse (Fig.~\ref{fig:emc_signal_params:}) contains $64$ samples and extends over a number of $T_\mathrm{b}$. It is parameterized with a function $U(t)$:
\begin{equation}\label{sigFunc}
U(t) = A \cdot F(t - \tau) + P,
\end{equation} 
where $A$ is the signal amplitude, $F(t)$ is the reference signal in the corresponding channel, $\tau$ is the time shift between reference and measured signals, $P$ is the signal pedestal.  The reference signal may vary from one EMC channel to another in shape and a peak position. A dedicated calibration procedure is used to retrieve reference signals for all EMC channels \cite{shapeCalib2019}. It is performed regularly using large data sets of $e^{+}e^{-} \to e^{+}e^{-}$  events. The reference signal is the averaged signal for  $e^{+}e^{-} \to e^{+}e^{-}$  events without a pedestal ($P = 0$) and with an amplitude $A = 1$ at a peak position $\tau_{ref}$. The signal shape is considered to be the same for all events due to relatively small light collection time inside counters.

\subsection{FLT event time shift}	
\label{subsec:FLTshift}
Since beams collide each $T_\mathrm{b}$ ns $\tau$ values are distributed close to collision times $T_{coll}^{n} = n\, T_\mathrm{b}$ where $n \in \mathbb{Z}$. Due to the current system design the FLT signal arrives at the right phase of the machine clock in $\sim90\%$ of events. Hence, the majority of signals from these events has $\tau$ values concentrated near $T_{coll}^{0}$ ($0$~ns). But in $\sim10\%$ of events the majority of the EMC signals are shifted close to non-zero  $T_{coll}^{n}$ times. This can take place when the FLT signal is produced prematurely as a response for some machine background event that happened close enough in time. It could also happen  due to bad time resolution of the FLT signal when it arrives at the wrong phase of the machine clock. A FLT event time shift should be detected for such events, as this time parameter can be used to correct event reconstruction in DC (Sec.~\ref{subsec:DC_reconstruction}) and  for reducing "extra" photons (Sec.~\ref{timeFilter}).

The FLT event time shift is obtained using EMC data during signals processing as described below in Sec.~\ref{sec:processing} and can take one of $T_{coll}^{n}$ values.

\section{Signal processing}
\label{sec:processing}
The values of the signal parameters ($A, P, \tau$) are extracted by fitting the function $U(t)$ to the signal  using either of two available algorithms \cite{instr20}: a linearization regression algorithm or a correlation function algorithm. The linearization regression algorithm is used for the most of the EMC signals (described in \cite{LinAlgo2017}). The  correlation function algorithm uses a correlation function between  measured and reference signals to extract the time shift (described in Sec.~\ref{sec:corfit}). Both algorithms provide two additional results of the signal processing: a quality of the fit and a return code.

\begin{figure}[ht]
	\centering
	\includegraphics[width=0.9\columnwidth]{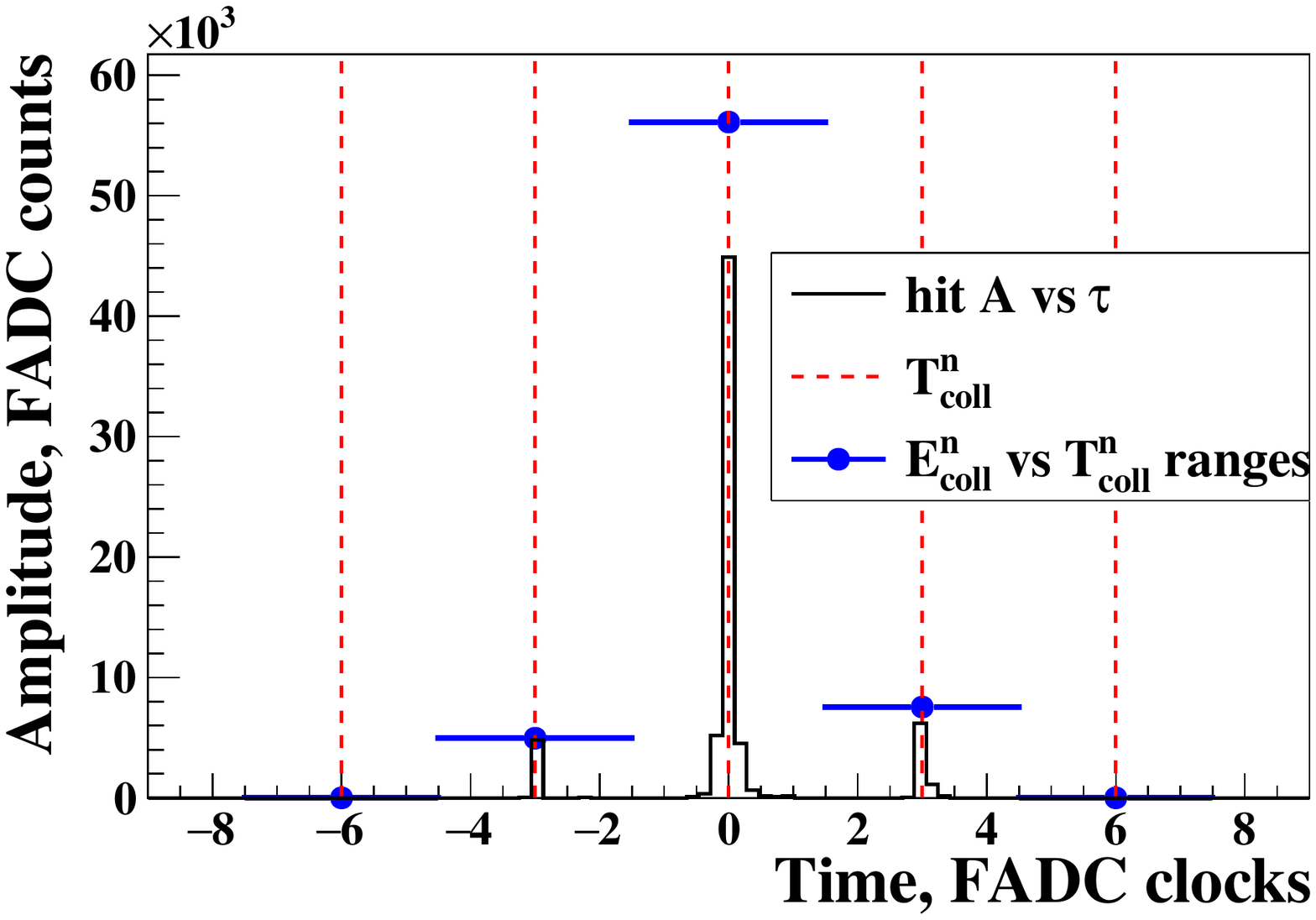}
	\protect\caption{The FLT event time shift calculation for one event. The histogram represents distribution of hits amplitudes over the hits time spectrum. Vertical dashed lines are collision times. Circles with horizontal bars represent total energy depositions $E_{coll}^{n}$ inside collision ranges with centers at $T_{coll}^{n}$ times.  $T_{coll}^{0}$ value is chosen as the FLT event time shift for this case.}
	\label{fig:time_shift_scheme:}
\end{figure}

Each event is processed in the following steps:
 \begin{enumerate}[Step 1.]
	\item  \textbf{Processing signals  with  estimated $A > 50$ FADC counts}. The linearization regression algorithm is used for the most of these signals. Special cases (strongly shifted and saturated  signals) are processed by the  correlation function algorithm.  Values of   $A, P, \tau$  are determined for each signal.
	\item \textbf{The FLT event time shift is calculated.}  $A,\tau$ values of all signals successfully processed at the previous step are used to determine from which beam collision the event takes origin.   The hit is considered to belong to the  collision of time $T_{coll}^{n}$ if the hit $\tau$ value lies inside its range $[(n - 0.5)\, T_\mathrm{b}, (n + 0.5)\, T_\mathrm{b}]$. In this way,  the closest $T_{coll}^{n}$ is determined for each hit. Then for each non-empty $T_{coll}^{n}$ range its total energy deposition  $E_{coll}^{n}$  is calculated as a sum of  $A$ values of the hits inside the range. The event time shift is determined as $T_{coll}^{n}$ time of the range with the biggest $E_{coll}^{n}$ (Fig.~\ref{fig:time_shift_scheme:}).
	\item \textbf{Processing signals with low energy deposition.} The rest of the signals are processed to extract only $A$ and $P$ parameters  using the linearization algorithm  with the signal time  $\tau$ being fixed at the FLT event time shift value. 
	\item \textbf{Calibration}. All values of $A$ and $\tau$ obtained at the previous steps are translated from FADC counts and clocks  to energy deposition in MeV and to time values in ns, respectively. 
\end{enumerate}

\subsection{The correlation function algorithm}
\label{sec:corfit}
\begin{figure*}
	\centering
	\includegraphics[width=0.9\columnwidth]{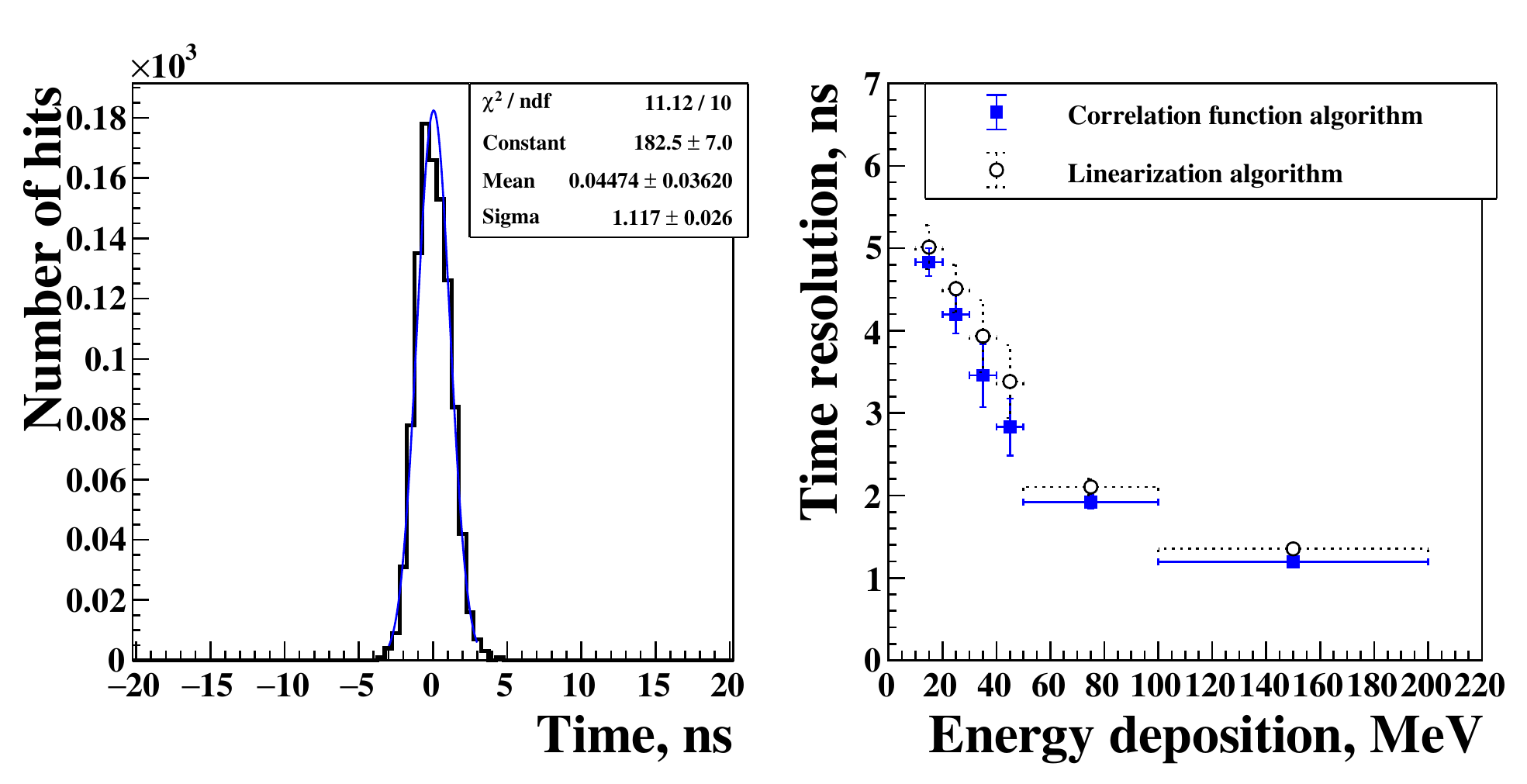}
	\protect\caption{The correlation function algorithm results on $e^{+}e^{-} \to e^{+}e^{-}$ events. Left: the obtained time resolution in one EMC channel for signals with  energy deposition $E > 100$~MeV. Right: the time resolution vs EMC energy deposition for one EMC channel in comparison with the results of the linearization algorithm.   
		\label{fig:corfuncres:}}
\end{figure*}

The correlation function algorithm  was implemented as an alternative for cross-checking the results of the linearization regression algorithm and for processing special cases like strongly shifted and saturated  signals. It processes a signal in two sequential steps: 
 \begin{enumerate}[Step 1.]
	\item The time $\tau$  is determined by finding the maximum of the correlation function  $\Omega(\tau)$ which is constructed  between the measured  and the reference signals as follows:
	\begin{equation}\label{eq:corfunc}
	\Omega(\tau) = \sum_{j=0}^{63} Y_{j} \cdot F_{j}(\tau),
	\end{equation} 
	where $j \in [0,63]$  is a signal sample number,  $Y_{j} = U(t_{j}) \,-\ P_{calib}$ is a $j$-th measured signal sample amplitude without a calibrated pedestal for the EMC channel, $F_{j} = F(t_{j} \,-\, \tau)$ is the reference signal amplitude calculated at the $j$-th sample time with respect to $\tau$ (the time shift between signals). 
	The maximum is obtained as a minimum of $-\Omega(\tau)$. The minimization is performed using the GNU Scientific Library (GSL) \cite{GSL_manual} implementation of the Brent's algorithm and GSL fast Fourier transform (FFT) methods. The first guess for the time shift $\tau_0$ is calculated  using the  Fourier transform shift property:
	\begin{equation}\label{fisrtGuess}
	\widehat{Y_j}(\omega) = e^{-i\omega\tau_0}\cdot \widehat{F_j}(\omega),\quad  \omega = \frac{2\pi}{64}, 
	\end{equation} 
	where $\widehat{Y_j}(\omega)$ and $\widehat{F_j}(\omega)$ are the output of the discrete Fourier transform for ${Y_j}$ and ${F_j}$ respectively. The  $\tau_0$ calculation is performed in two iterations. The discrete nature of the transform leads to differential non-linearity.
	\item The values of $A$ and $P$ are extracted using the linear regression model, where the obtained $\tau$ value is applied to shift $F(t)$ along the time axis.
\end{enumerate}	
The obtained time resolution for one EMC channel and its dependence on EMC energy deposition is almost the same as we achieved with the linearization algorithm (Fig.~\ref{fig:corfuncres:}). This correlation function algorithm can successfully process almost all signals, but it is relatively  slow (up to $7-10$ times). 

\begin{figure*}
	\centering
	\includegraphics[width=0.9\columnwidth]{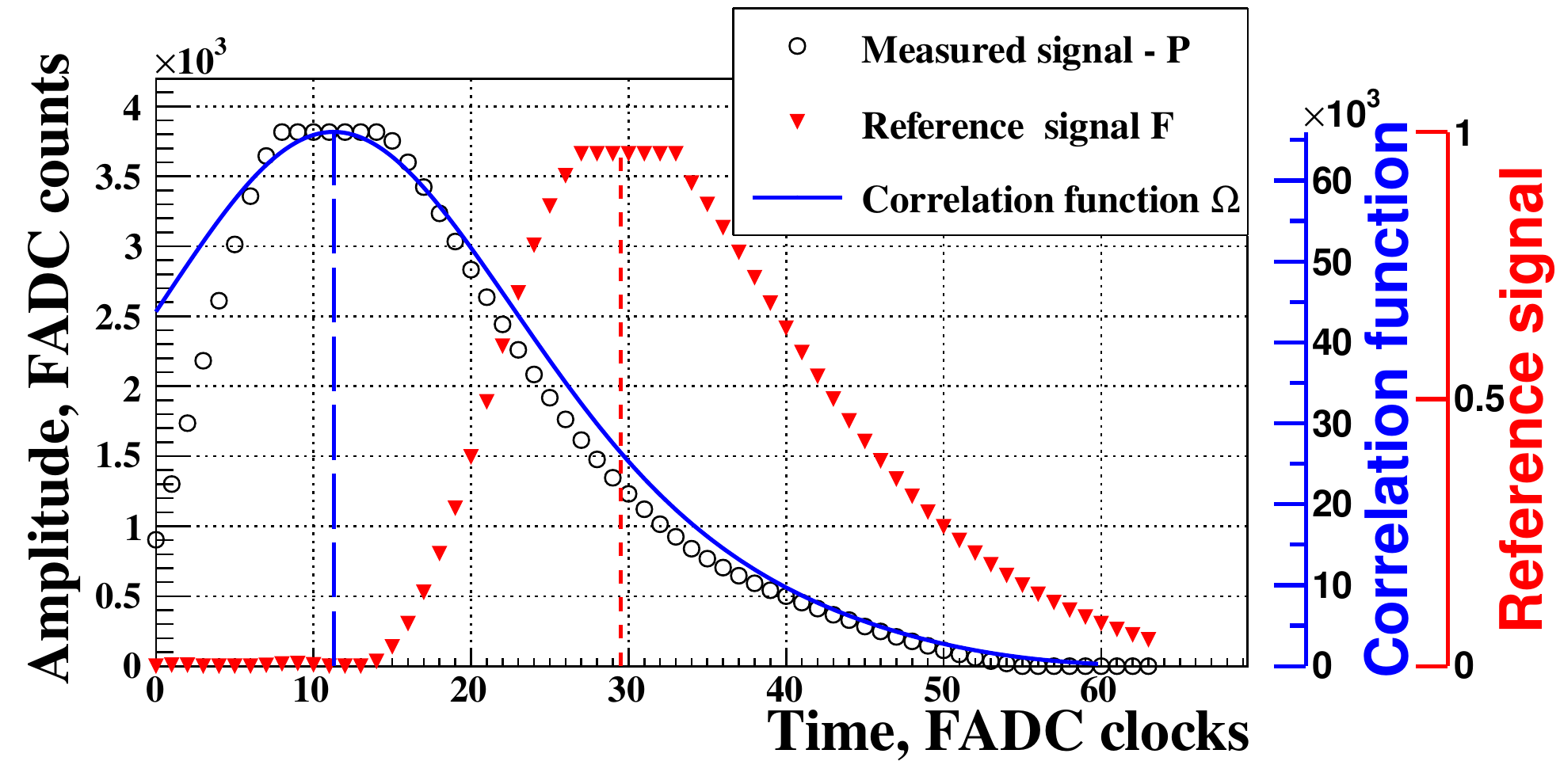}
	\protect\caption{Step 1 of the correlation function algorithm for a saturated signal. Empty-circle (black) points are the measured signal without $P$. Reversed (red) triangles are  the transformed reference signal shape with  imitated  saturation of the measured signal used  for $\tau_0$ calculations. The obtained $\tau_0$ value is applied as the first guess for the $-\Omega(\tau)$ minimization. Solid (blue) line represents the correlation function $\Omega(\tau)$ that reaches its maximum at the point $\tau + \tau_{ref}$. 
		\label{fig:corfitcase:}}
\end{figure*}

We apply the algorithm for processing saturated and strongly shifted signals. The EMC signal is considered to be saturated if more than one signal sample has full amplitude $> 4094$~FADC clocks. Shifted signals  reach maximum  at more than $7$~FADC clocks away from $t_{ref}$ and can occur as a result of previous collision events, cosmic-ray events, machine background pileup or from nuclear interactions with matter. The algorithm was tuned for these cases in the following ways:
\begin{itemize}
	\item  To process saturated signals, we made changes in both algorithm steps. To calculate the FFT first guess $\tau_0$ we use a transformed version of the reference signal shape $F(t)$. The transformation imitate the saturation of $F(t)$ on a signal height, where $F(t)$ signal has the same width as the saturation width of the measured signal. The transformed in this way $F(t)$  and the resulted correlation function for a saturated signal are visualized in Fig.~\ref{fig:corfitcase:}. At the second step, in which values for $A$ and $P$ are determined, the saturated signal samples of the measured and transformed reference signals are not taken into account. Here only  the signal amplitude  $A$ is determined with $P$ being fixed to the calibrated value.
	\item To process strongly shifted signals only $A$ is calculated at the algorithm second step with $P$ being fixed to the calibrated value.  
\end{itemize}
The algorithm results for these cases have been validated on MC signals with known properties \cite{instr20}.
\section{Simulation}
\subsection{EMC channel response simulation}
\label{subsec:MC}
EMC simulation software has been modified to produce realistic response of the EMC  electronics. 
The Geant4 framework \cite{Geant4} is used to simulate particle passage through matter in our simulation software. Each EMC crystal is divided into several sensitive cells. When an EMC crystal is triggered by a  simulated particle passing through it, several cell hits are generated for this crystal by the Geant4 core. Energy deposition and time from all crystal cell hits are used to generate EMC channel response. The resulted signal sample $U_{i}$ at each sampling time $t_i$ is calculated as follows:
\begin{equation}\label{eq:simFuncc}
	U_{i} = k_\mathrm{chan} \cdot \sum_{j}{ (E_j \cdot F(t_{i} \,-\, t_j \,-\, t_\mathrm{shift}))\,+\, P \,+\, P_\mathrm{noise}},
\end{equation} 
where $k_\mathrm{chan}$ is the conversion coefficient from  MeV to FADC counts in the channel, $j$ is the cell hit number, $E_j, t_j$ is the energy deposition and  time of the cell hit $j$ respectively, $t_\mathrm{shift}$ is the common time shift that can be set, $P$ is the channel pedestal calibrated value, $P_\mathrm{noise}$ is the pedestal noise generated for all samples $U_i$ using the channel calibrated noise. The cell hit time is calculated relatively to the event start time as:
\begin{equation}\label{simFuncCellTime}
 t_j = (t_\mathrm{prestep} \:+\:  t_\mathrm{poststep}) \: / \: 2 \: + \: t_\mathrm{flight},
\end{equation} 
where $t_\mathrm{prestep}$ and $t_\mathrm{poststep}$ are Geant4 core pre-step and post-step times respectively,  $t_\mathrm{flight}$ is the light time of flight in a straight line from the cell center to the phototriod plane. 

Since MC signals are generated using the reference signal obtained on $e^{+}e^{-} \to e^{+}e^{-}$ events the processing procedure described in Sec.~\ref{sec:processing}  gives delayed time results for MC signals. To avoid this, we calibrate time delays for $e^{+}e^{-} \to e^{+}e^{-}$ MC signals for each EMC layer, since  these events should have $\tau$ values near zero by construction of the measurement procedure. The obtained time delays are applied in generation of all EMC pulses as a $t_\mathrm{shift}$ from Eq.~\ref{eq:simFuncc}.

\subsection{Pileup and the time channel}
High energy physics experiments have a well known problem of a high background noise. This effect is needed to be reproduced in the simulation for the purposes of physics analysis. To do that, we collect and store background (BKG) events during data taking. BKG events are events triggered by a fixed frequency signal synchronized with the machine clock.  These events are superimposed with Geant4 events resulting in more realistic simulation data ("mixed data"). This means that hits from MC data are "mixed" with hits from  BKG events forming from two sets of hits the resulted one for each subsystem.

After the EMC electronics upgrade the measured time is also taken into account for mixing EMC hits. Two versions of the procedure have been implemented: simplified mixing and mixing with superimposing signal pulses.

The simplified  mixing procedure uses the values of signal parameters after the calibration step of the processing procedure (Sec.~\ref{sec:processing}). If there is either MC or BKG hit in the event for the given EMC channel, the hit is taken as it is into the resulted set of hits. If there are both types of the hits in the event for the EMC channel, the procedure takes the hit with the larger energy deposition. 

The second version of the mixing procedure is  a more complicated case, since it works with signal pulses before extracting the values of signal parameters.  To superimpose signals correctly, it should be taken into account, that MC and  BKG signals can have different calibration coefficients used to transform energy deposition from FADC counts into MeV. To avoid incorrect reconstructed particle energies BKG signal pulses are scaled to match MC calibration settings. 

Furthermore, a small part of the BKG signals can be time shifted on one FADC clock ($T_\mathrm{s}$)  with respect to  $T_{coll}^{n}$ times because of the phase shift between the machine clock and the generator trigger signal, which took place at the start of using the upgraded EMC electronics. To avoid incorrect superimposing on MC signals, these BKG pulses are transformed by shifting them back. In case of negative BKG signal time shift a pedestal sample is added  at the start of the signal and the last sample is truncated. In case of positive time shift the first sample is truncated and the last sample of the original signal is duplicated. This electronics issue has been fixed and synchronization stability has been put under constant monitoring.

\begin{figure*}
	\centering
	\includegraphics[width=0.49\columnwidth]{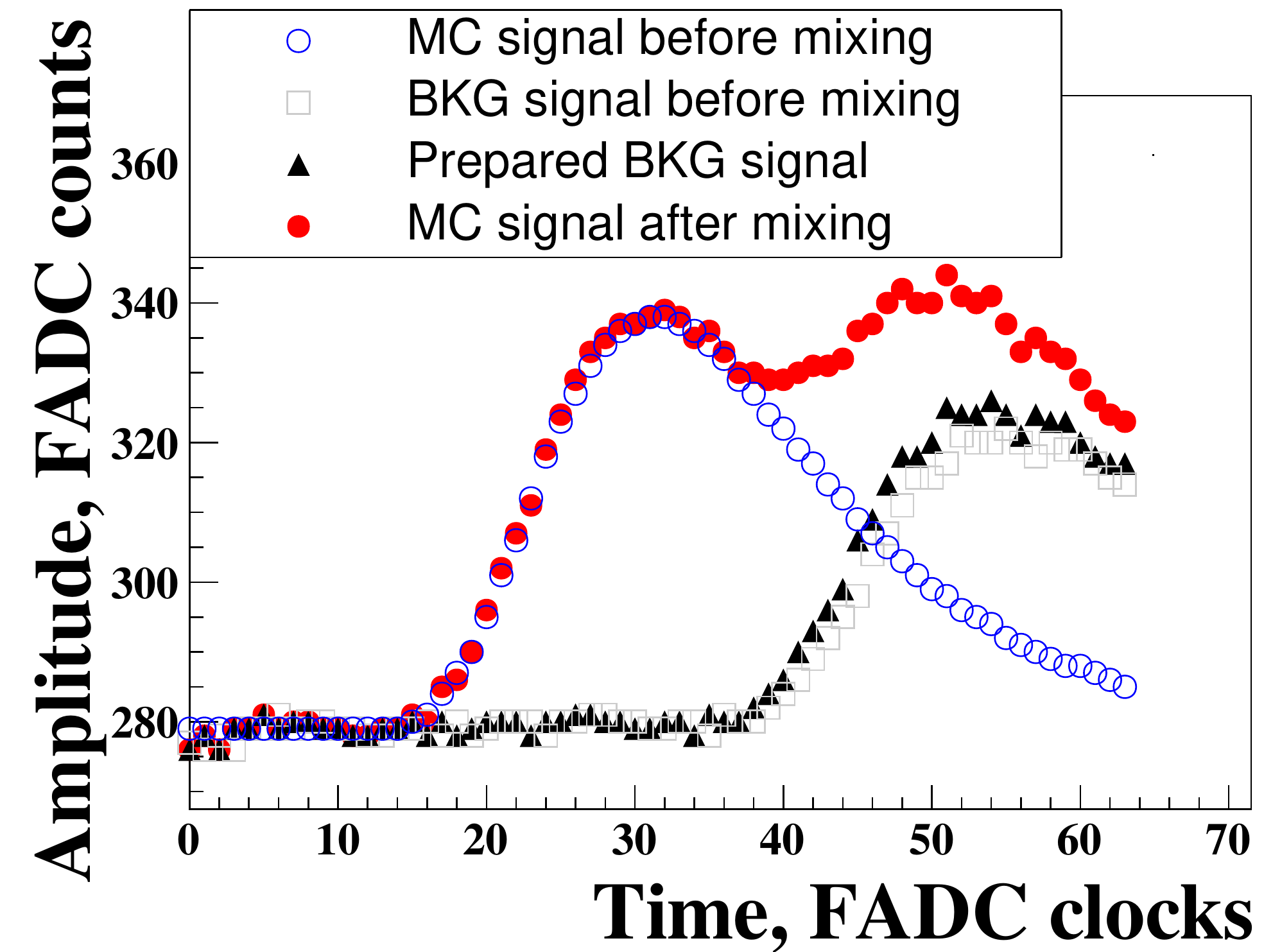}
	\includegraphics[width=0.49\columnwidth]{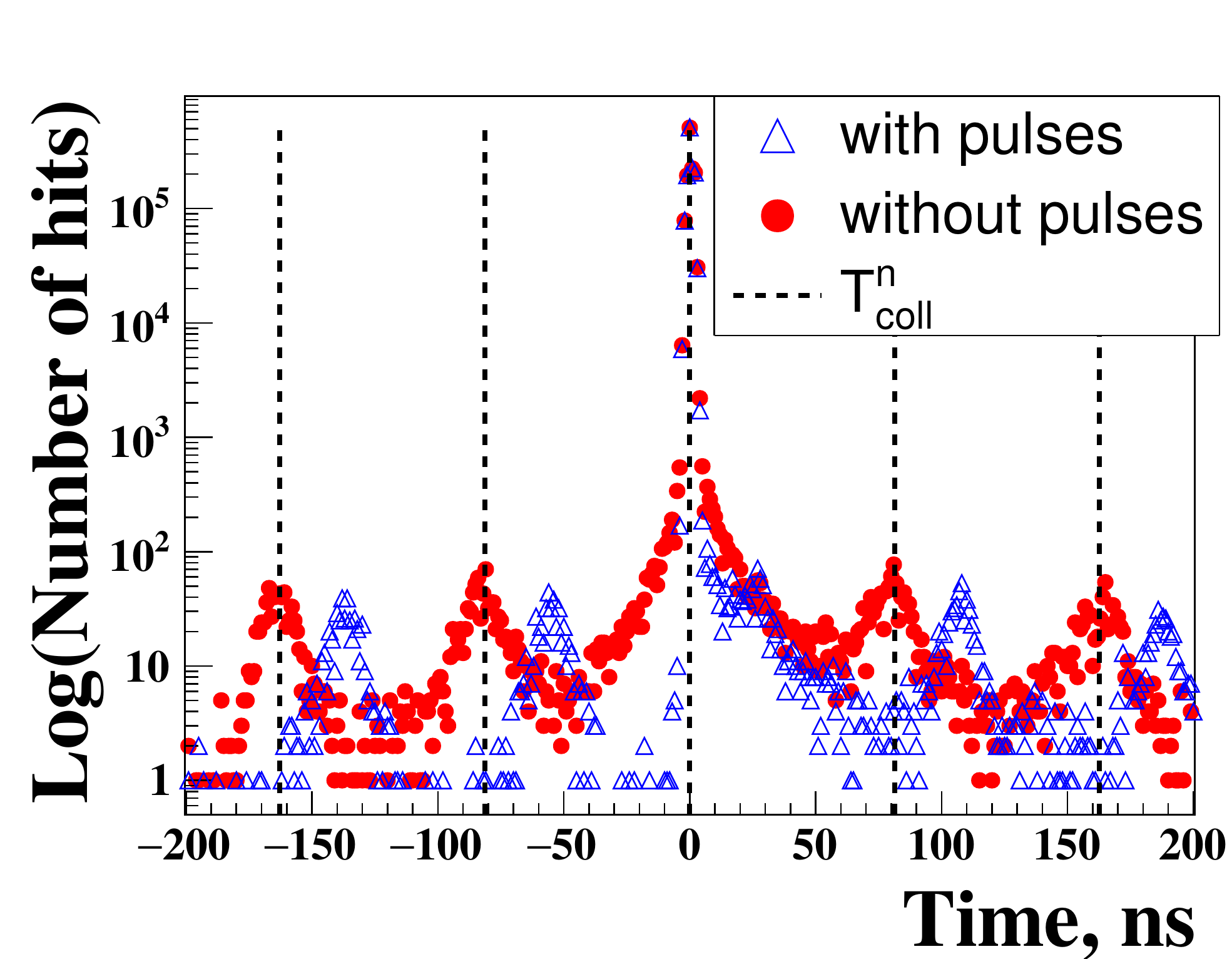}
	\protect\caption{The results of mixing EMC signal pulses.  Left: an illustrative example of MC signal pulse  before and after superimposing a prepared (scaled and shifted in time) BKG signal pulse.  Right: hits time spectrum for   $e^{+}e^{-} \to e^{+}e^{-}$ events from MC data  with  the simplified mixing procedure (blue triangles) and the procedure  superimposing EMC signal pulses (red circles). Vertical dashed lines illustrate beam collision times $T_{coll}^{n}$. 
		\label{fig:mixpics:}}
\end{figure*}

An illustrative example of how the MC signal pulse changes after superimposing a BKG signal is presented in  Fig.~\ref{fig:mixpics:} (left).  Time spectra of $\tau$ values for two versions of the  mixing procedure are presented on Fig.~\ref{fig:mixpics:} (right). The spectrum of the simplified procedure has time peaks shifted relative to  $T_{coll}^{n}$ times. But the spectrum of the superimposing procedure has correct peaks near $T_{coll}^{n}$ times due to correctly prepared BKG signal pulses before  mixing.

\section{Time measurement applications}
\label{sec:timeusages}
\subsection{Event reconstruction}
Event reconstruction with EMC time  in the SND offline framework allows to obtain several important time parameters: an EMC particle time, FLT and EMC event time shifts. The FLT event time shift reconstruction is discussed in Sec.~\ref{subsec:FLTshift} and Sec.~\ref{sec:processing}, others are described below.

\subsubsection{Reconstructed particle time}
Particle time calculation uses results of  an EMC clusterization procedure, which constructs clusters from hits taking into account only a hit energy deposition and where it happened in space. The resulted clusters participate in forming particle candidates. Each such cluster contains data about its  hits. Successfully extracted $\tau$ values of cluster hits are used to calculate particle  minimal ($t_{min}$) and mean ($t_{mean}$) times:
\begin{equation}\label{partForms}
t_{min} = \min\limits_{i} \tau_i,\ t_{mean} = \frac{\sum_{i}\tau_i \cdot \sigma(E_i)}{\sum_{i} \sigma(E_i)},
\end{equation} 
where $i$ is a cluster hit index, $\tau_i$ is a cluster hit time, $\sigma(E_i)$ is an EMC channel time resolution that depends on hit energy deposition $E_i$. These time parameters are also calculated for each EMC layer separately over particle hits. 

\begin{figure}[ht]
	\centering
	\includegraphics[width=0.9\columnwidth]{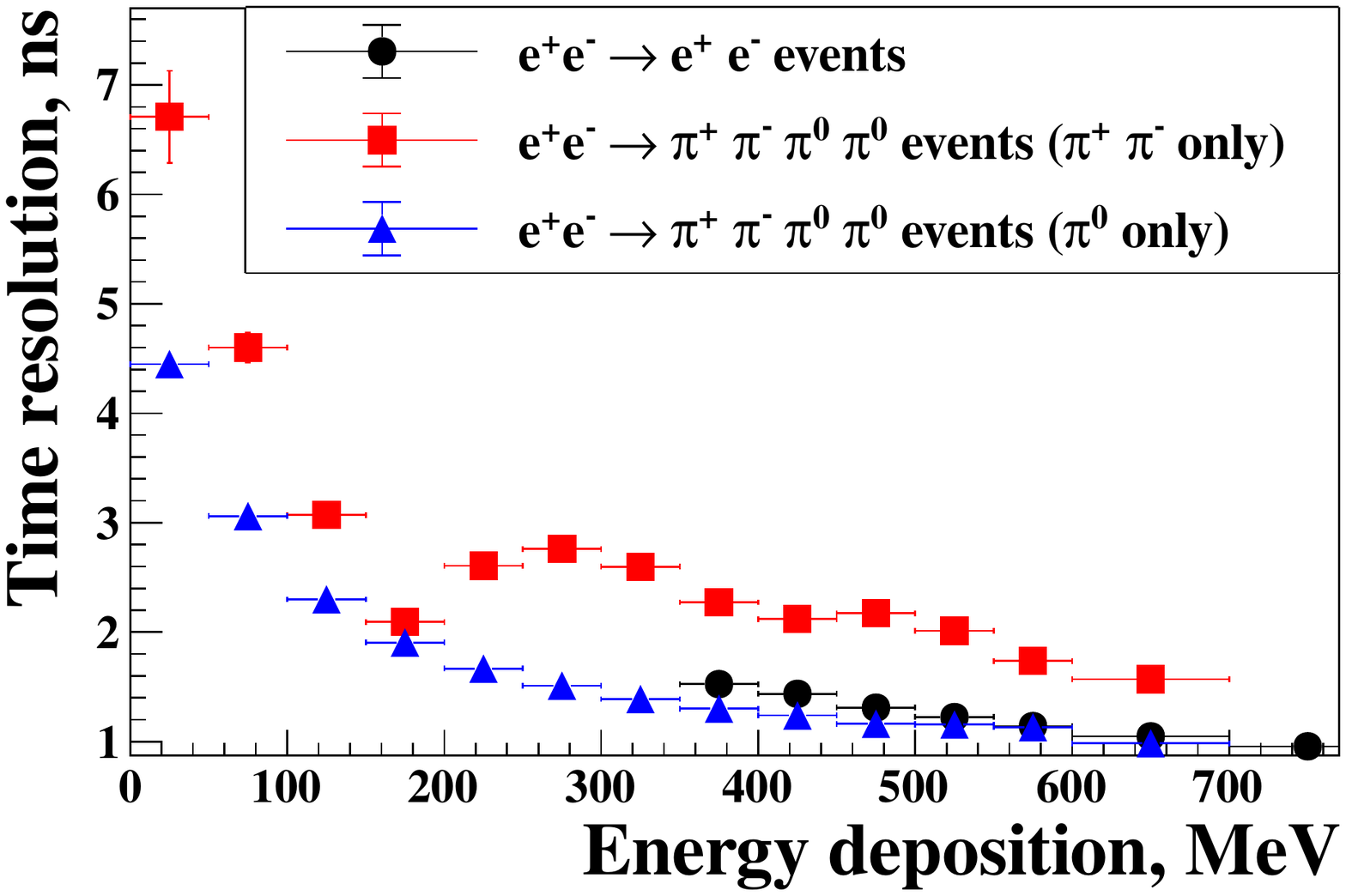}
	\protect\caption{Mean particle time resolution vs  particle energy deposition for $e^{+}e^{-} \to e^{+}e^{-}$  and $e^{+}e^{-} \to \pi^{+}\pi^{-}\pi^{0}\pi^{0}$ events. }
	\label{fig:partpics:}
\end{figure}

In Fig.~\ref{fig:partpics:}  mean  time resolution is presented for particles from  $e^{+}e^{-} \to e^{+}e^{-}$  and $e^{+}e^{-} \to \pi^{+}\pi^{-}\pi^{0}\pi^{0}$ events in several particle energy deposition ranges. The time resolution differs for particle types because they have different energy distribution inside the EMC. The behavior of the resolution for charged pions can be explained by nuclear interaction and particle decay inside the EMC volume, which can produce delayed hits.

\subsubsection{Drift time correction in tracks reconstruction}
\label{subsec:DC_reconstruction}
\begin{figure}[ht]
 	\centering
 	\includegraphics[width=0.9\columnwidth]{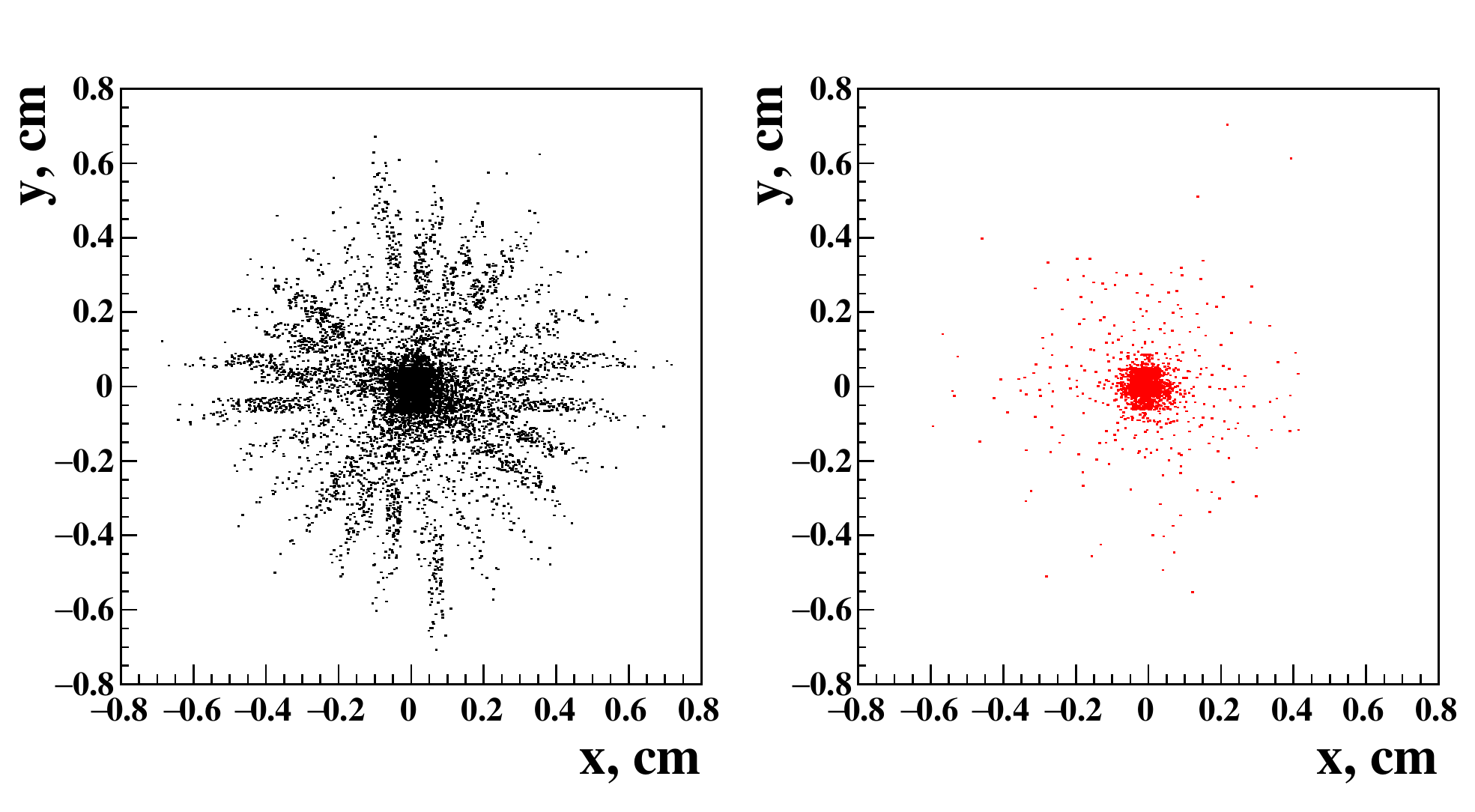}
 	\protect\caption{Track parameters for $e^{+}e^{-} \to e^{+}e^{-}$  events with non-zero FLT time shift without (left) and with (right) drift time correction.}
 	\label{fig:dcpics:}
 \end{figure}

The track coordinates along the direction of the ionization drift in a DC cell are calculated using the ionization drift time measurements. These time measurements are performed relative to the FLT signal, which is synchronized with the machine clock as  described in Sec.~\ref{sec:channel}. If the FLT event time shift is non-zero (see Sec.~\ref{subsec:FLTshift}) the calculations of the track parameters are corrected to take the shift into account. Track parameters from events with a non-zero FLT time shift are presented in Fig.~\ref{fig:dcpics:}. Without the time shift correction DC signals concentrate at DC cell borders.  

\subsubsection{EMC event time shift}
\label{subsec:emshift}
For physics analysis the EMC event time shift is provided in several forms. The first one indicates the FLT time shift and can take only one of $T_{coll}^{n}$ values as described in Sec.~\ref{subsec:FLTshift}. But there are also signals from  processes with delayed detection by the SND subsystems ($e^{+}e^{-} \to n\bar{n}$, $e^{+}e^{-} \to p\bar{p}$) and pileups from hits unrelated to the FLT signal. To separate these signals the EMC event time shift is also calculated  without strict binding to $T_{coll}^{n}$ times. 

For  $e^{+}e^{-} \to n\bar{n}$ analysis \cite{nan2022}  it is obtained as an average EMC time over hits  from  a time cluster with the largest total energy deposition in the event:
\begin{equation}\label{eq:nanEvTime}
t_{nan} = \frac{\sum_{i}\tau_i \cdot E_i}{\sum_{i} E_i},
\end{equation} 
where  $i$ is the cluster hit index, $\tau_i$ is the cluster hit time, $E_i$ is the cluster hit energy deposition. To perform  time clusterization all hits are sorted by time first. Then a hit is added to the cluster if it  has time difference less then $10$~ns with the closest in time cluster hit.
\subsection{Background suppression}
Background events can be of cosmic-ray or beam-induced origin.  

\begin{figure}[ht]
	\centering
	\includegraphics[width=0.49\columnwidth]{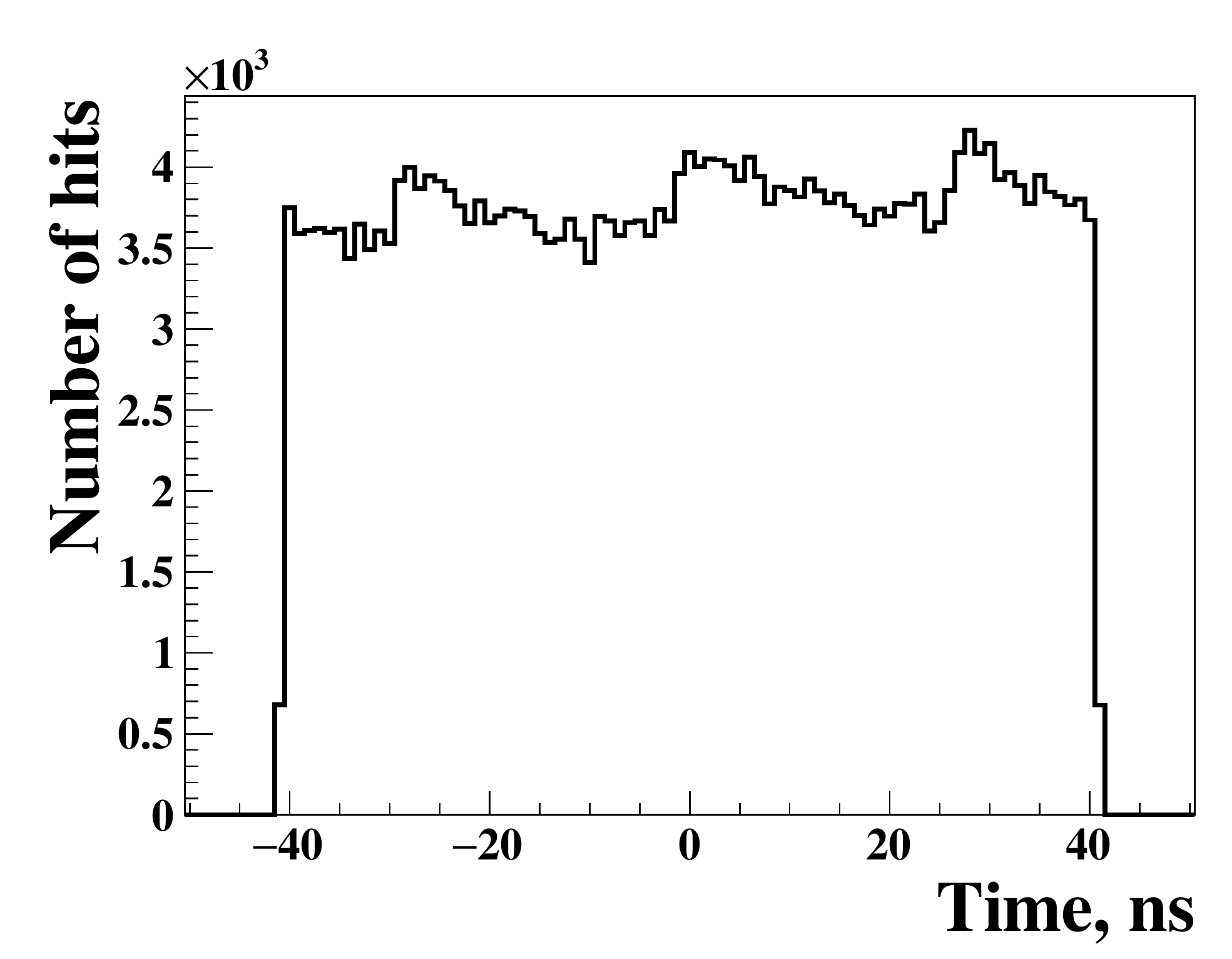}
	\includegraphics[width=0.49\columnwidth]{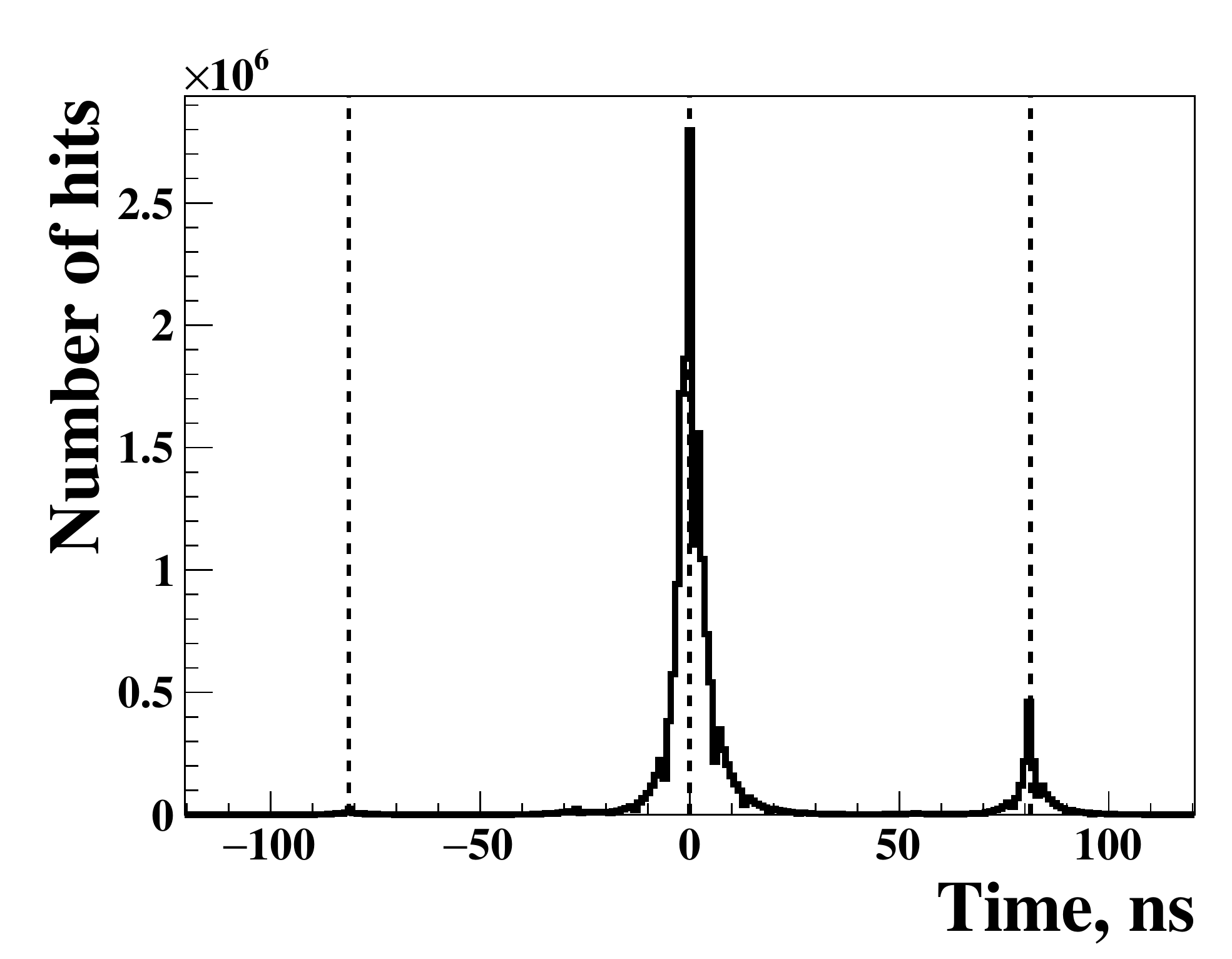}
	\protect\caption{EMC hits time spectra. Left: cosmic-ray events. Right: the events without signals in TS, which were selected with the SND muon veto; vertical lines represent beam collision times $T_{coll}^{n}$. \label{fig:timespectrum:}}
\end{figure}

The cosmic-ray background is usually suppressed using a muon veto. However, the muon detector in not $100\%$ efficient  due to its design limitations. EMC time can be used as an additional tool since cosmic-ray events are evenly distributed in time. The EMC hits time distribution for cosmic-ray events is presented in Fig.~\ref{fig:timespectrum:} (left). Small nonuniformity of the distribution arise from differential nonlinearity of the correlation function algorithm used to obtain this spectrum (Sec.~\ref{sec:corfit}). 

The beam-induced background can produce EMC hits with  time  close to beam collision times $T_{coll}^{n}$. But unlike these  hits collision event hits concentrate mostly inside the FLT event time shift range. This can be used to minimize beam-induced background effects like "extra" photons.  

Fig.~\ref{fig:timespectrum:} (right)  illustrates time properties of background and collision events mentioned above.

\subsubsection{Reducing "extra" photons}
\label{timeFilter}
The EMC signals from the machine-induced background can lead to reconstruction of "extra" photons. A special procedure has been implemented to detect such artifacts. An EMC hit is removed from the EMC clusterization if its time parameter was successfully retrieved during signal processing, and its value differs from the FLT event time shift on more than $1.5\,T_\mathrm{b}$. 

After applying the procedure on recorded data \cite{TIPP21}, the number of "extra" photons decreases by $5.6\%$ for $e^{+}e^{-} \to \gamma\gamma$ events with two or more reconstructed EMC clusters ($n_{\mathrm{c}}$) and the number of these events decreases by $0.1\%$. That also leads to the increase  in the number of $e^{+}e^{-} \to \gamma\gamma$ events with strictly $n_\mathrm{c}=2$ by  $0.2\%$. The number of $e^{+}e^{-} \to e^{+}e^{-}$ events with $n_\mathrm{c}=2$ increases by $0.2\%$.

\subsubsection{\texorpdfstring{$e^{+}e^{-} \to n\bar{n}$}{e+e- -> nantin} analysis}
\label{nan}
The $e^{+}e^{-} \to n\bar{n}$ events represent a specific $e^{+}e^{-}$ annihilation process since they can be recognized only by antineutron annihilation inside the EMC through detection of large and strongly asymmetric energy deposition without corresponding charged tracks. 

The suppression of cosmic-ray background  is an important task in the  $n\bar{n}$ analysis \cite{nan2022}. The number of cosmic-ray and  $e^{+}e^{-} \to n\bar{n}$ events are almost equal after applying basic criteria with the muon veto for event selection. However, EMC time spectrum can be used for further suppression since events of different types have different time distribution. 

Antineutrons travel with low velocity therefore EMC  $e^{+}e^{-} \to n\bar{n}$ signals have EMC times which are delayed  in comparison with signals from typical  $e^{+}e^{-}$ annihilation processes like for example,  $e^{+}e^{-} \to \gamma\gamma$. On the other hand, cosmic-ray events are uniformly distributed, while the EMC times of the residual beam background events are close to the beam collision times  $T_{coll}^{n}$. 
\begin{figure*}[ht]
	\centering
	\includegraphics[width=0.8\columnwidth]{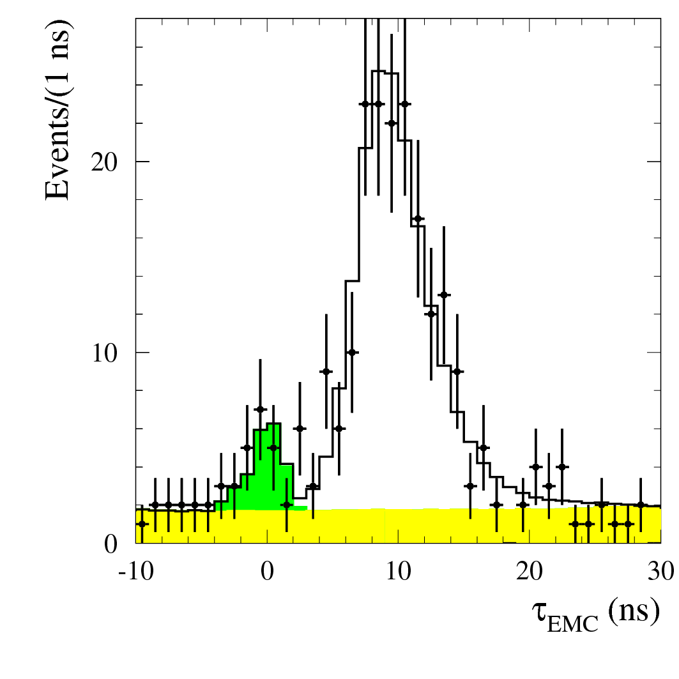}\hspace{1pc}
	\caption{The distribution of event averaged EMC time for the selected $n\bar{n}$ events in 2019 season at the beam energy $945$~MeV \cite{nan2022}. Points with error bars represent data. The solid histogram is the result of the fit described in the text: the light-shaded (yellow) histogram region is the fitted cosmic-ray background, the medium-shaded (green) region is the fitted beam-induced and physical background. 
		\label{fig:nan_pics:}}
\end{figure*}
To extract the number of $e^{+}e^{-} \to n\bar{n}$ events the  spectrum of an average EMC event time  is constructed. The average event time is calculated according to Eq.~\ref{eq:nanEvTime}. The  spectrum for  the $n\bar{n}$ analysis of the 2019 season data at the beam energy $945$~MeV \cite{nan2022} is presented in Fig.~\ref{fig:nan_pics:}. Here $n\bar{n}$ signals produce a clearly distinguishable wide delayed peak near $10$~ns, beam background gives the small peak at the beam collision time $T_{coll}^{0}$, while the uniform component of the spectrum  represents cosmic-ray background.

\section{Summary}
The upgraded EMC spectrometric channel of the SND detector provides digitized signal pulses, which are processed by two algorithms to extract signals amplitudes and arrival times. The linearization regression algorithm processes most of the signals, while the correlation function algorithm, described in detail, is applied only for processing  strongly shifted and saturated pulses.  The time resolution in the EMC channel is $\sim1$~ns for energy deposition over $100$~MeV. New functionality of the SND software for the EMC channel response simulation is described. It enables generating EMC digitized signal pulses, which are then used together with real background signals to imitate pileup. Several time parameters are calculated during event reconstruction: a particle time, FLT and  EMC event time shifts. The EMC measured time applications are discussed and presented. The FLT time shift  helps to correctly reconstruct DC track parameters in and suppress the "extra" photons from the beam-induced background. The possibilities of EMC time measurements are used in the $e^{+}e^{-} \to n\bar{n}$ analysis \cite{nan2022}.   

\section*{Acknowledgments}
The work was supported by RSF grant 23-22-00011 (Sec.~\ref{subsec:emshift}, Sec.~\ref{nan}) and by Ministry of science and higher education of the Russian Federation (the rest of the work).
\bibliography{article_bib}
\end{document}